\documentclass[conference]{IEEEtran}
\usepackage[cmex10]{amsmath}
\usepackage{graphicx}
\usepackage{epstopdf}
\usepackage{cite}
\usepackage{amsfonts}
\usepackage{algorithm} 
\usepackage{algorithmic} 
\usepackage[letterpaper, left=1in, right=1in, bottom=1in, top=0.75in]{geometry}

\begin{document}
\title{Energy Efficiency Maximization for CoMP Joint Transmission with Non-ideal Power Amplifiers}
\author{\IEEEauthorblockN{Yuhao~Zhang, Qimei~Cui, and Ning~Wang}
\IEEEauthorblockA{School of Information and Communication Engineering,\\
Beijing University of Posts and Telecommunications, Beijing, 100876, China\\
Email: cuiqimei@bupt.edu.cn}}

\maketitle

\begin{abstract}
Coordinated multipoint~(CoMP) joint transmission~(JT) can save a great deal of energy especially for \mbox{cell-edge} users due to strengthened received signal, but at the cost of deploying and coordinating cooperative nodes, which degrades energy efficiency~(EE), particularly when considerable amount of energy is consumed by non-ideal hardware circuit. In this paper, we study energy-efficient cooperation establishment, including cooperative nodes selection~(CNS) and power allocation, to satisfy a required data rate in coherent JT-CoMP networks with non-ideal power amplifiers~(PAs) and circuit power. The selection priority lemma is proved first, and then the formulated discrete combinatorial EE optimization is resolved by proposing node selection criterion and deriving closed-form expressions of optimal transmission power. Therefore, an efficient algorithm is provided and its superiority is validated by Monte Carlo simulations, which also show the effects of non-ideal PA and the data rate demand on EE and optimal number of active nodes.
\end{abstract}

\vspace{2mm}

\section{Introduction}
Recently, the energy efficiency~(EE) of wireless networks has drawn much research attention because energy consumption is growing rapidly and will soon reach intolerable levels with the evolution of information and communication industry~\cite{Ref1,Ref6}. Through joint and coordinated schedule, the coordinated multipoint~(CoMP) transmission is a technology designed to increase cell coverage and improve spectral efficiency~(SE) of wireless networks or, alternatively, save notable amounts of energy, especially when serving \mbox{cell-edge} users~\cite{Ref5,Ref3}. Joint transmission CoMP~(JT-CoMP), one of the most promising CoMP techniques, can be characterized by simultaneous transmission from multiple cooperative nodes to a single user with fully data and control information exchange~\cite{Ref11}. As a result, it can achieve better performance compared with other technologies. There are two methods for JT-CoMP, i.e., coherent and non-coherent transmission, where the coherent JT-CoMP is more attractive since it reaps coherent combining gain that strengthens received signal quality as well as decreases spatial interference~\cite{Ref8}. Therefore, we mainly study the optimal cooperation establishment, including cooperative nodes selection~(CNS) and power allocation, to maximize the EE in coherent JT-CoMP networks.

In practice, the CNS is the key and fundamental problem that needs to be resolved firstly when designing the cooperation establishment scheme. It is known that when interference created outside the cluster of cooperative nodes is neglected, the SE maximization is obtained when the number of cooperative nodes is as high as possible. And when \mbox{out-of-cluster} interference is considered, the capacity of the \mbox{JT-CoMP} networks saturates after surpassing a maximum number of cooperative nodes~\cite{Ref4}. To the best of our knowledge, the EE optimal CNS in JT-CoMP networks has not been studied properly so far, not to mention the coherent transmission scheme.

For a single user, more cooperative nodes involved in JT-CoMP can reduce transmission power due to larger received signal power and less interference~\cite{Ref4}, but at the cost of deploying more hardware infrastructures and coordinating these nodes, which both cause more energy consumption~\cite{Ref12}. This cost becomes more serious in practice, where the power amplifiers~(PAs) are always non-ideal, whose efficiency varies nonlinearly along with the output transmission power~\cite{PC_model}. For example, quite large power consumption independent with the transmission power will be consumed by envelope-tracking PA~(ETPA)~\cite{Reference43,Ref7}. Therefore, the EE of the system can be improved significantly if the best trade-off between its benefit and cost is achieved. Our main objective is to study the EE optimal resource allocation with consideration of this trade-off, which has never been discussed in the existing literatures.

In this paper, we propose energy-efficient CNS and power allocation in coherent JT-CoMP networks satisfying a required data rate with non-ideal PAs and circuit power. The EE maximization, which turns out to be a discrete and continuous mixed combinatorial optimization, is resolved in three steps, corresponding to the proposed CNS and power allocation~(CNS-PA) scheme. Simulations are conducted to validate the superiority of our proposed scheme as well as the effect of non-ideal PAs. Our main contributions are summarized as follows.
\begin{itemize}
  \item Considering ETPA and circuit power, the joint CNS and power allocation scheme is proposed to optimize the EE of coherent JT-CoMP networks.
  \item The selection priority lemma is proved and the closed-form expressions of optimal transmission powers are derived, based on which the node selection criterion is demonstrated to maximize EE.
  \item Simulation results verify the superiority of the proposed scheme and reveal that less cooperative nodes can obtain better EE under ETPA and circuit power.
\end{itemize}

\section{System Model and Problem Formulation}\label{Sec2}
In the considered JT-CoMP networks, single-antenna transmission nodes~(TNs) are independently distributed over the researched rectangular area $(D_1 \times D_2)$ according to a two-dimensional spatial homogeneous Poisson point process~(PPP) with density of $\zeta$. Without loss of generality, a common single-antenna receiver, e.g., the user equipment~(UE), located at the origin $(0,0) \in \mathbb{R}^2$ is studied, which is assumed to be served by $M$ cooperative TNs, denoted by $\mathcal{M}$, being able to provide the strongest signal power. With a required data rate $R_{\rm{dl}}$, the $M$ TNs transmit the same information symbols $s$ towards the common receiver in every periodical time frame duration~$T$. Moreover, these $M$ cooperative TNs can share all the information and are connected through high-speed and low-latency backhaul links, e.g., optical fibers, with perfect time-frequency synchronization.

Considering large scale path loss and Rayleigh fading, the block fading channel model is used to characterize the complex channel gain between TN~$m$ and the receiver, denoted by $h_m$. It is assumed that the channel information is detected and estimated at the receiver perfectly and then transmitted back to its corresponding TN. Therefore, the real-time channel state information~(CSI), including both amplitude $\{{\left| {{h_m}} \right|}\}$ and phase information $\{e^{j \angle{h_m}}\}$, is available in \mbox{JT-CoMP} transmission, which together with data can be exchanged through backhaul links among $\mathcal{M}$ without error and delay. $W$ is the system wireless bandwidth and the additive white Gaussian noise~(AWGN) $n$ at the receiver has zero mean and variance $P_N = N_0 W$, where $N_0$ is the power spectral density~(PSD) of the noise.
\subsection{System Capacity}\label{Sec2:1}
Within the time frame duration $T$, the desired information $s$ is transmitted to the common receiver by $\mathcal{M}$ with transmission power $P_m(\geq 0)$ respectively. It is noted that $P_m=0$ means TN~$m$ is in idle mode and do not take part in transmission. Therefore, the received signal at the receiver can be expressed as
\begin{equation}\label{ReceSignal}
    y=\sum\limits_{m=1}^{M}{\sqrt{{{P}_{m}}}{{h}_{m}}x_m}+i_{\rm{out}}+n,
\end{equation}
where $x_m = (w_m \, s)$ is the copy of the information symbol $s$ that TN~$m$ transmits using the weighting factor $w_m$. Note that $\mathbb{E}\left\{ {{{\left| {{s}} \right|}^2}} \right\} = 1$. And $i_{\rm{out}}$ is \mbox{out-of-cluster} interference with power $I_{\rm{out}}$, which is created by other TNs outside $\mathcal{M}$. For simplicity, $i_{\rm{out}}$ is modeled by AWGN for average performance.

In this paper, coherent transmission scheme is considered, which needs amplitude and phase information related to the channel on all cooperative TNs. In the coherent \mbox{JT-CoMP} networks, the phase compensation is made first before joint transmission since the ideal real-time CSI can be obtained. Therefore, the weighting factor $w_m = e^{-j \angle{h_m} }$ for coherent \mbox{JT-CoMP} scheme, based on which the achievable downlink data rate of the receiver can be given by
\begin{equation}\label{coh_AchiRate}
    r_{\rm{dl}} = W \cdot {\log _2}\Big[{1 + \frac{\Big({\sum\limits_{m = 1}^M {\sqrt {{P_m}} {\left| {{h_m}} \right|}}}\Big)^2}{I_{\rm{out}}+P_N}}\Big].
\end{equation}
\subsection{Practical Energy Consumption}\label{Sec2:2}
ETPA is considered in this paper, since it will introduce quite large energy consumption independent with the transmission power, which deteriorates the EE significantly~\cite{Reference43,Ref7}. The total power consumption at TN~$m$ can be presented by~\cite{EEdelay}
\begin{equation}\label{ETPA}
    \Psi_{\rm{ETPA}}(P_m) = \frac{P_m + a P_{\max,m}}{(1+a) \eta_{\max,m}},
\end{equation}
where $P_{\max,m}$ and $\eta_{\max,m}$ are maximum output power and maximum PA efficiency of TN~$m$, respectively~\cite{Reference43,Ref7}. For simplicity, it is assumed the ETPA equipped at all TNs are the same and with identical parameters. It is clear that the ideal power amplifier~(IPA) is a special case of ETPA by letting $a = 0$.

The circuit power consumption can be further decomposed into static and dynamic components. The static component $P_{\rm{base}}$ is constant and depends on the hardware (to drive hardware), whereas the dynamic component $P_{{c}} = \varepsilon \cdot r_{\rm{dl}}$ represents the power consumption for signal processing blocks, e.g., analog and digital signal processing, and depends on the actual downlink data rate $r_{\rm{dl}}$ where coefficient $\varepsilon$ is the power for transmitting a data bit~\cite{EnergyCom}. Therefore, the total power consumption for transmission of TN~$m$ can be given by
\begin{equation}\label{ptx}
    P_{{\rm{tx}},m}=\Psi_{\rm{ETPA}}(P_m) + \varepsilon \cdot r_{\rm{dl}} + P_{{\rm{base}},{\rm{tx}}}.
\end{equation}

Similarly, the power consumption for reception can be formulated, as given by
\begin{equation}\label{prx}
    P_{\rm{rx}}=\varepsilon \cdot r_{\rm{dl}} + P_{{\rm{base}},{\rm{rx}}}.
\end{equation}
The power consumption in idle mode, denoted by $P_{\rm{idle}}$, is assumed to be constant, i.e., independent of $r_{\rm{dl}}$.
\subsection{Optimization Problem Formulation}\label{Sec2:3}
The EE is defined as the ratio between the number of overall data bits transmitted and the total energy consumed by all nodes, denoted by $E_{\rm{total}}$, within the duration $T$~\cite{Reference43,Ref7}, as given by
\begin{equation}\label{EE_indicator}
{\eta _E} = \frac{{R_{\rm{dl}} \cdot T}}{{{E_{\rm{total}}}}} = \frac{{R_{\rm{dl}} \cdot T}}{{{P_{\rm{total}}} \cdot T}} = \frac{R_{\rm{dl}}}{P_{\rm{total}}},
\end{equation}
which indicates that given the required downlink data rate $R_{\rm{dl}}$, maximizing $\eta_E$ is equivalent to minimizing the total power consumption, denoted by $P_{\rm{total}}$.

It is assumed that there are $\bar{M} (\leq M)$ active TNs and the other $(M-\bar{M})$ TNs are in idle mode. Therefore, the EE maximization problem can be formulated as
\begin{align}\label{G_EE_Opt}\tag{\textbf{P1}}
    \mathop {\min }\limits_{{\{P_m\}},\bar{M}} \quad & \sum\limits_{m = 1}^{\bar{M}} {P_{{\rm{tx}},m}} + (M-\bar{M}) {P_{\rm{idle}}} + {P_{{\rm{rx}}}}, \\
    {\rm{s.t.}} \quad & {R_{\rm{dl}}} \le {r_{\rm{dl}}}, \nonumber \\
    & \bar{M} \in \{1, 2, \cdots, M\}, \nonumber \\
    & 0 \le \Psi \left( {{P_m}} \right) \le {P_{\max,m}},\ m \in \{1, 2, \cdots, \bar{M}\}. \nonumber
\end{align}

For optimal solution of~\eqref{G_EE_Opt}, ${r_{\rm{dl}}} = {R_{\rm{dl}}}$ is achieved by reducing transmission powers $\{P_m\}$ until the data rate constraint is taken. It is clear that the maximum achievable data rate $R_{\rm{max}}^{\bar{M}}$ is reached only when all the $\bar{M}$ active TNs transmit with the maximum power. If $R_{\rm{max}}^{\bar{M}}<R_{\rm{dl}}$, there is no feasible solution, i.e., the following constraint must be satisfied, as given by
\begin{equation}\nonumber
	(2^{\frac{R_{\rm{dl}}}{W}} -1) \leq {\frac{{{\Big( {\sum\limits_{m = 1}^{\bar{M}} {\sqrt {{\Psi ^{ - 1}}\left( {{P_{\max ,m}}} \right)} {{\left| {{h_m}} \right|}}} } \Big)^2}}}{I_{\rm{out}}+P_N}},
\end{equation}
where ${\Psi ^{ - 1}}$ denotes the inverse function of~\eqref{ETPA}.

To solve problem~\eqref{G_EE_Opt}, we need to find out the optimal number of active TNs, denoted by $\bar{M}^*$, and figure out which $\bar{M}^*$ TNs are selected, as well as their optimal transmission powers $P_m^*$. It is obvious that~\eqref{G_EE_Opt} is a discrete combinatorial optimization but even with continuous optimization variable, which is hard and nontrivial to resolve by the standard optimization methods.
\section{EE Optimization under ETPA}\label{Sec3}
In this section, we solve the nontrivial combinatorial optimization problem~(\ref{G_EE_Opt}) under ETPA, which can be divided into three subproblems.
\subsection{The Selection Priority}\label{Sec3:1}
Selecting optimal active TNs from the cooperative cluster relies on the channel conditions and their impacts on the total power consumption in our considered situation. It is obvious that there is a selection priority for different TNs, where the better TN has a higher priority to be selected. It is seen that only the channel coefficients are different among these $M$ TNs due to the same ETPA parameters we assumed, based on which the following lemma about the selection priority is presented as

\vspace{2mm}

\newtheorem{lemma}{\textbf{Lemma}}
\begin{lemma} \label{Selection_Priority}
\textit{Under ETPA and circuit power, the activated $\bar{M}^*(\leq M)$ TNs have the better channel coefficients ${\left| {{h_m}} \right|} ,\ m \in \{1, 2, \cdots, \bar{M}^*\}$, compared with other idle $(M-\bar{M}^*)$ TNs for the optimal EE, as given by}
\begin{equation}\nonumber
\min \{\left| {{h_i}} \right|,1 \hspace{-0.5mm} \leq \hspace{-0.5mm} i \hspace{-0.5mm} \leq \hspace{-0.5mm} \bar{M}^*\} \geq \max \{\left| {{h_j}} \right|,\bar{M}^* \hspace{-1mm} +1 \hspace{-0.5mm} \leq \hspace{-0.5mm} j \hspace{-0.5mm} \leq \hspace{-0.5mm} M \}.
\end{equation}
\end{lemma}


\begin{IEEEproof}
Firstly, it is assumed that this lemma does not establish and there is a optimal selection result, called Type~I, where a TN whose channel coefficient is better than one active TN is in idle mode. Without loss of generality, the former TN is assumed as TN $i \ (i \in \{1, 2, \cdots, \bar{M}^*\})$ and the latter one is TN $j \ (j \in \{\bar{M}^*+1, \bar{M}^*+2, \cdots, M\})$, where ${\left| {{h_j}} \right|}>{\left| {{h_i}} \right|}$. Then, another selection result, called Type~II, can be constructed by replacing TN $i$ with TN $j$, while other $(\bar{M}^*-1)$ active TNs remain unchanged. According to our assumption, Type~I is more energy-efficient than Type~II.

Considering the constraint of fixed required data rate, Type~I and Type~II must reach the same $R_{\rm{dl}}$, based on which, according to \eqref{coh_AchiRate}, the following relationship must be satisfied, as given by
\begin{equation}\nonumber
	\sum\limits_{m = 1, m \neq i}^{\bar{M}^*} \hspace{-4mm} {\sqrt {{P_m}} {\left| {{h_m}} \right|}} + \sqrt{P_i} {\left| {{h_i}} \right|} = \hspace{-4mm} \sum\limits_{m = 1, m \neq i}^{\bar{M}^*} \hspace{-4mm} {\sqrt {{P_m}} {\left| {{h_m}} \right|}} + \sqrt{P_j} {\left| {{h_j}} \right|}.
\end{equation}

It can be obtained that $P_j<P_i$ due to ${\left| {{h_j}} \right|}>{\left| {{h_i}} \right|}$, therefore, Type~II consumes less energy according to the objective function in problem~\eqref{G_EE_Opt}, i.e., Type~II is more energy-efficient than Type~I, which is contradict to our first assumption; in other words, the opposite of our assumption is established, i.e., the optimal $\bar{M}^*$ active TNs have the best channel coefficients among the cooperative cluster.
\end{IEEEproof}

\vspace{2mm}

\textbf{Remark:} The lemma can resolve the subproblem of which TNs should be involved in the joint transmission for optimal EE if the optimal $\bar{M}^*$ is known.

\vspace{2mm}

According to {Lemma}~\ref{Selection_Priority}, the optimal $\bar{M}^*$ TNs have the best ${\left| {{h_m}} \right|} ,\ m \in \{1, 2, \cdots, \bar{M}^*\}$ compared with other idle TNs. Therefore, we sort the $M$ TNs among the cooperative cluster in descending order of ${\left| {{h_m}} \right|}$, denoted by $\mathbf{S}$, which can be given by
\begin{equation}\nonumber
{\left| {{h_1}} \right|} \geq {\left| {{h_2}} \right|} \geq {\left| {{h_3}} \right|} \geq \cdots \geq {\left| {{h_{\bar{M}^*}}} \right|} \geq \cdots \geq {\left| {{h_M}} \right|},
\end{equation}
and the active $\bar{M}^*$ TNs are the first $\bar{M}^*$ TNs in $\mathbf{S}$.

Following {Lemma}~\ref{Selection_Priority}, we can formulate the optimal transmission power for all active TNs for any given $\bar{M}$, where Lemma~\ref{Selection_Priority} is ensured for the TNs selection priority; in other words, $\bar{M}$ TNs with larger channel coefficients are selected, while others are in idle mode.
\subsection{The Optimal Power Allocation}\label{Sec3:2}
In this subsection, given a certain $\bar{M}$, we explore the subproblem of the optimal transmission power allocation for the involved TNs in JT-CoMP networks. By substituting \eqref{ETPA} and \eqref{ptx} into \eqref{G_EE_Opt}, the objective function $E_{\rm ETPA}$ of problem~\eqref{G_EE_Opt} can be formulated, as given by
\begin{equation}\label{EE_Opt_ETPA}
\begin{aligned}
E_{\rm{ETPA}} & = \sum\limits_{m = 1}^{\bar{M}} {\frac{{{P_m} + a {P_{\max,m}}}}{{\left( {1 + a} \right) {\eta _{\max,m}}}}}  + \bar{M} \, P_{{\rm{base}},{\rm{tx}}} \\
& + (M-\bar{M}) P_{\rm{idle}} + \varepsilon \cdot R_{\rm{dl}} + P_{{\rm{rx}}}.
\end{aligned}
\end{equation}

For this fixed case of the original problem~\eqref{G_EE_Opt}, the EE optimal power allocation for these $\bar{M}$ TNs can be obtained, as presented in the following theorem.

\vspace{2mm}

\newtheorem{theorem}{\textbf{Theorem}}
\begin{theorem} \label{PowerA_ETPA_con}
\textit{Under ETPA and circuit power, for optimal EE,given a certain $\bar{M}$, the optimal transmission power of TN $m \ (m \in \{1, 2, \cdots, \bar{M}\})$ must meet the following formula, as given by}
\begin{equation}\label{Pm_ETPA}
    {P_m^*} = \frac{({2^{\frac{R_{\rm{dl}}}{W}}-1})\frac{\eta _{\max,m}^2{\left| {{h_m}} \right|}^2}{{I_{\rm{out}}+P_N}}}{(\sum\limits_{m = 1}^{\bar{M}}\frac{\eta _{\max,m}{\left| {{h_m}} \right|}^2}{{I_{\rm{out}}+P_N}})^2}.
\end{equation}
\end{theorem}


\begin{IEEEproof}
The Lagrange function that combines the objective function and data rate constraint of~\eqref{G_EE_Opt} is
\begin{equation}
    {\mathcal{L}_{\rm{ETPA}}} = {E_{\rm{ETPA}}} - {\lambda _{\rm{ETPA}}}{{({\sum\limits_{m = 1}^{\bar{M}} {\sqrt {{P_m}} {{\left| {{h_m}} \right|}}} })^2}},
\end{equation}
where ${\lambda _{\rm{ETPA}}}$ is the Lagrange multiplier. According to \mbox{Karush-Kuhn-Tucker}~(KKT) conditions, \mbox{$\frac{{\partial^2 {\mathcal{L}_{\rm{ETPA}}}}}{{\partial {P_m}^2}} = 0$} should hold simultaneously for all active TN~$m$ in any optimal solution of the problem. Therefore, the following relationship must be verified for all $m$, as given by
\begin{equation}\label{Pm_B_ETPA}
    {P_m^*} = {B_{\rm{ETPA}}} \, \eta _{\max,m}^2 \, {{\left| {{h_m}} \right|}^2},
\end{equation}
where
\begin{equation}
    {B_{\rm{ETPA}}} = \lambda_{\rm{ETPA}}^2{\left( {1 + a} \right)^2} ({I_{\rm{out}}+P_N}) (2^{\frac{R_{\rm{dl}}}{W}}-1).
\end{equation}
By substituting~\eqref{Pm_B_ETPA} into the data rate constraint ${r_{\rm{dl}}} = {R_{\rm{dl}}}$, ${B_{\rm{ETPA}}}$ can be resolved consequently, as given by
\begin{equation}\label{BETPA}
    {B_{\rm{ETPA}}} = \frac{({2^{\frac{R_{\rm{dl}}}{W}}-1})}{({I_{\rm{out}}+P_N})(\sum\limits_{m = 1}^{\bar{M}}\frac{\eta _{\max,m}{\left| {{h_m}} \right|}^2}{{I_{\rm{out}}+P_N}})^2}.
\end{equation}
Combining \eqref{Pm_B_ETPA} and \eqref{BETPA}, it is clear that ${P_m}>0$ if only $R_{\rm{dl}}>0$. Moreover, the optimal transmission power $P_m^*$ at TN $m$ is proved to be the form of \eqref{Pm_ETPA}.
\end{IEEEproof}

\vspace{2mm}

\textbf{Remark:} The theorem can illustrate the optimal transmission powers of the given $\bar{M}$ TNs involved in joint transmission in coherent JT-CoMP networks, based on which a new EE optimization problem with optimal transmission power allocation can be formulated.

\vspace{2mm}

By substituting \eqref{Pm_ETPA} into \eqref{EE_Opt_ETPA}, the new reformulated EE optimization problem, only depending on the number of active TNs $\bar{M}$, can be obtained, as given by
\begin{align}\label{Re_Opt_ETPA_con}\tag{\textbf{P2}}
    \mathop {\min }\limits_{\bar{M}} \quad & \alpha_{\bar{M}} ({2^{\frac{R_{\rm{dl}}}{W}}} \hspace{-0.5mm} - \hspace{-0.5mm} 1) \hspace{-0.5mm} + \hspace{-1mm} \sum\limits_{m = 1}^{\bar{M}} \hspace{-0.5mm} {\frac{{a {P_{\max,m}}}}{{\left( {1 + a} \right) {\eta _{\max,m}}}}} \hspace{-0.5mm} + \hspace{-0.5mm} \beta_{\bar{M}}, \\
    {\rm{s.t.}} \quad & \bar{M} \in \{1, 2, \cdots, M\}, \nonumber
\end{align}
where
\begin{equation}\nonumber
    \alpha_{\bar{M}} = \frac{1}{{\left( {1 + a} \right)\sum\limits_{m = 1}^{\bar{M}} {\frac{{\eta _{\max,m}}{{{\left| {{h_m}} \right|}^2}}}{{{I_{\rm{out}}+P_N}}}} }},
\end{equation}
\begin{equation}\nonumber
    \beta_{\bar{M}} = \bar{M} \, P_{{\rm{base}},{\rm{tx}}} + (M - \bar{M}) P_{\rm{idle}} + \varepsilon \cdot R_{\rm{dl}} + P_{{\rm{rx}}}.
\end{equation}

Given $\bar{M}$, the overall power consumption $E_{\rm{ETPA}}^{\bar{M}}$ can be calculated via the objective function of \eqref{Re_Opt_ETPA_con}.
\subsection{The Cooperative TNs Establishment}\label{Sec3:3}
It can be seen that the reformulated problem~\eqref{Re_Opt_ETPA_con} is a discrete optimization where the objective function is different for every $\bar{M}$. In order to find the optimal solution, we need to calculate the overall power consumption $E_{\rm{ETPA}}^{\bar{M}}$ for all possible $\bar{M}$, which is very complicated and impractical, especially when $M$ is large. Moreover, it is obvious that the optimal number of active TNs $\bar{M}^*$ has no closed-form expression. Therefore, the extra TN selection criterion is introduced here for any $\bar{M}$ to accomplish the cooperative TNs establishment easily, which is described as follows.

\vspace{2mm}

\begin{theorem} \label{criterion_joining}
\textit{Under ETPA and circuit power, given $\bar{M}$, the additional TN $(\bar{M}+1)$ will be activated for optimal EE if and only if the following condition satisfies}
\begin{equation}\label{Judge_joining}
    \frac{\Gamma(\bar{M}+1)}{\sum\limits_{m = 1}^{\bar{M}} \Gamma(m) \cdot \sum\limits_{m = 1}^{\bar{M}+1} \Gamma(m) }>\frac{\theta_{\bar{M}+1}(1+a)}{2^{\frac{R_{\rm{dl}}}{W}}-1},
\end{equation}
\textit{where}
\begin{equation}\nonumber
    \Gamma(m) = {\frac{{\eta _{\max,m}{{\left| {{h_m}} \right|}^2}}}{{{I_{\rm{out}}+P_N}}}},
\end{equation}
\begin{equation}\nonumber
    \theta_{\bar{M}+1}={\frac{{a {P_{\max,\bar{M}+1}}}}{{\left( {1 + a} \right) {\eta _{\max,\bar{M}+1}}}}} + P_{{\rm{base}},{\rm{tx}}}-P_{\rm{idle}}.
\end{equation}
\end{theorem}

\vspace{3mm}

\begin{IEEEproof}
For a given $\bar{M}$, the extra TN $(\bar{M}+1)$ will be divided into active TNs for energy saving if $E_{\rm{ETPA}}^{\bar{M}+1}<E_{\rm{ETPA}}^{\bar{M}}$, i.e., $E_{\rm{ETPA}}^{\bar{M}+1}-E_{\rm{ETPA}}^{\bar{M}}<0$, which can be reformulated as
\begin{equation}\nonumber
    \left(\alpha_{\bar{M}+1}-\alpha_{\bar{M}}\right)({2^{\frac{R_{\rm{dl}}}{W}}} - 1) + \theta_{\bar{M}+1}< 0.
\end{equation}

And then \eqref{Judge_joining} can be obtained through mathematic simplification and transformation.
\end{IEEEproof}

\vspace{2mm}

\textbf{Remark:} The theorem can determine how many TNs are in active mode for achieving optimal EE in coherent JT-CoMP networks.

\vspace{2mm}

Applying {Lemma}~\ref{Selection_Priority} and {Theorem}~\ref{criterion_joining}, the cooperative TNs establishment can be achieved by several judgments of~\eqref{Judge_joining}. Specifically, with the help of {Lemma}~\ref{Selection_Priority}, sequence $\mathbf{S}$ with descending order of all the $M$ TNs is first obtained, based on which the EE optimal active TNs can be then established by using {Theorem}~\ref{criterion_joining}, where the optimal active TNs are the first $\bar{M}^*$ TNs in sequence $\mathbf{S}$.
\section{Algorithmic Implementation}\label{Sec4}
In our considered situation, the EE optimal cooperation establishment can be achieved by utilizing {Lemma}~\ref{Selection_Priority}, {Theorems}~\ref{PowerA_ETPA_con} and~\ref{criterion_joining}. Firstly, descending sort the $M$ TNs according to their channel coefficients ${\left| {{h_m}} \right|}$ and get a sequence of these $M$ TNs. Then let $\bar{M}=1$ and traverse the obtained sequence one by one to check whether~\eqref{Judge_joining} is satisfied. If it is satisfied, $\bar{M}\Leftarrow \bar{M}+1$ and continue this process, otherwise end this process and let $\bar{M}^*=\bar{M}$. After this checking process, the optimal active TNs are the first $\bar{M}^*$ TNs in the sequence. Finally, applying {Theorem}~\ref{PowerA_ETPA_con}, the optimal transmission powers of these active TNs can be formulated via~\eqref{Pm_ETPA}. Therefore, the CNS-PA algorithm is proposed, as summarized in {Algorithm~\ref{alg1}}, which has linear computational complexity in terms of $M$, i.e., $\mathcal{O}(M)$.
\begin{algorithm}[htbp]
\caption{CNS-PA algorithm}
\label{alg1}
\begin{algorithmic}[1]
\STATE Estimate all channel conditions and feedback $h_m$;
\STATE Descending sort the $M$ TNs according to ${\left| {{h_m}} \right|}$ to form a sequence of these $M$ TNs. And then $\bar{M} \Leftarrow 1$;
\STATE Check whether \eqref{Judge_joining} is satisfied; \label{judge_if}
\IF{\eqref{Judge_joining} is satisfied}
    \STATE $\bar{M} \Leftarrow \bar{M}+1$ and turn to Step \ref{judge_if};
\ELSE
    \STATE $\bar{M}^* \Leftarrow \bar{M}$;
\ENDIF
\STATE Calculate the optimal transmission powers $P_m^*$ for these $\bar{M}^*$ TNs according to \eqref{Pm_ETPA};
\STATE The first $\bar{M}^*$ TNs in $\mathbf{S}$ transmit with power $P_m^*$, while other TNs are all in idle mode.
\end{algorithmic}
\end{algorithm}
\section{Simulation results}\label{Sec5}
In this section, Monte Carlo simulations are carried out to validate our proposed CNS-PA scheme in coherent JT-CoMP networks with ETPAs and circuit power. The simulation parameters are specified in TABLE~I with reference to~\cite{PC_model,Reference43,Ref7,EEdelay}. Besides the proposed CNS-PA scheme, some other schemes are simulated for comparison purpose, as described by
\begin{itemize}
  \item {All nodes scheme}: All TNs are in active mode with uniform power allocation.
  \item {All nodes and PA scheme}: All TNs are in active mode with optimal power allocation.
  \item {Single node scheme}: Only one TN with best channel condition is in active mode.
  \item {CNS scheme}: Select the optimal active TNs but with uniform power allocation.
\end{itemize}

\begin{table}[bp]\label{Table_Simulation}
\centering
\caption{Simulation Parameters}
\begin{tabular}{l|l}
\hline
\textbf{Parameters} & \textbf{Values} \\
\hline
System bandwidth ($W$) & 10\,MHz \\
Noise power spectral density ($N_0$) & --\,174\,dBm/Hz \\
Number of Overall TNs ($M$) & 16 \\
Average path loss ($\rm{L}$) & $103.8+21\log_{10}d$\,dB \\
The density of TNs ($\zeta$) & 50\,BS/$\rm km^2$ \\
Length of rectangular area ($D_1, D_2$) & $1, 1$\,km \\
Idle circuit power ($P_{{\rm{idle}}}$) & 10\,mW \\
Static circuit power ($P_{\rm{base}}$) & 50\,mW \\
Dynamic circuit factor ($\varepsilon$) & 2\,mW/Mbps \\
Maximum output power ($P_{\max,m}$) & 46\,dBm \\
Maximum PAs efficiency ($\eta_{\max,m}$) & 0.35\\
Dependent parameter of ETPA ($a$) & 0.0082\\
\hline
\end{tabular}
\end{table}

\begin{figure}[!t]
\centering
\includegraphics[scale=0.25]{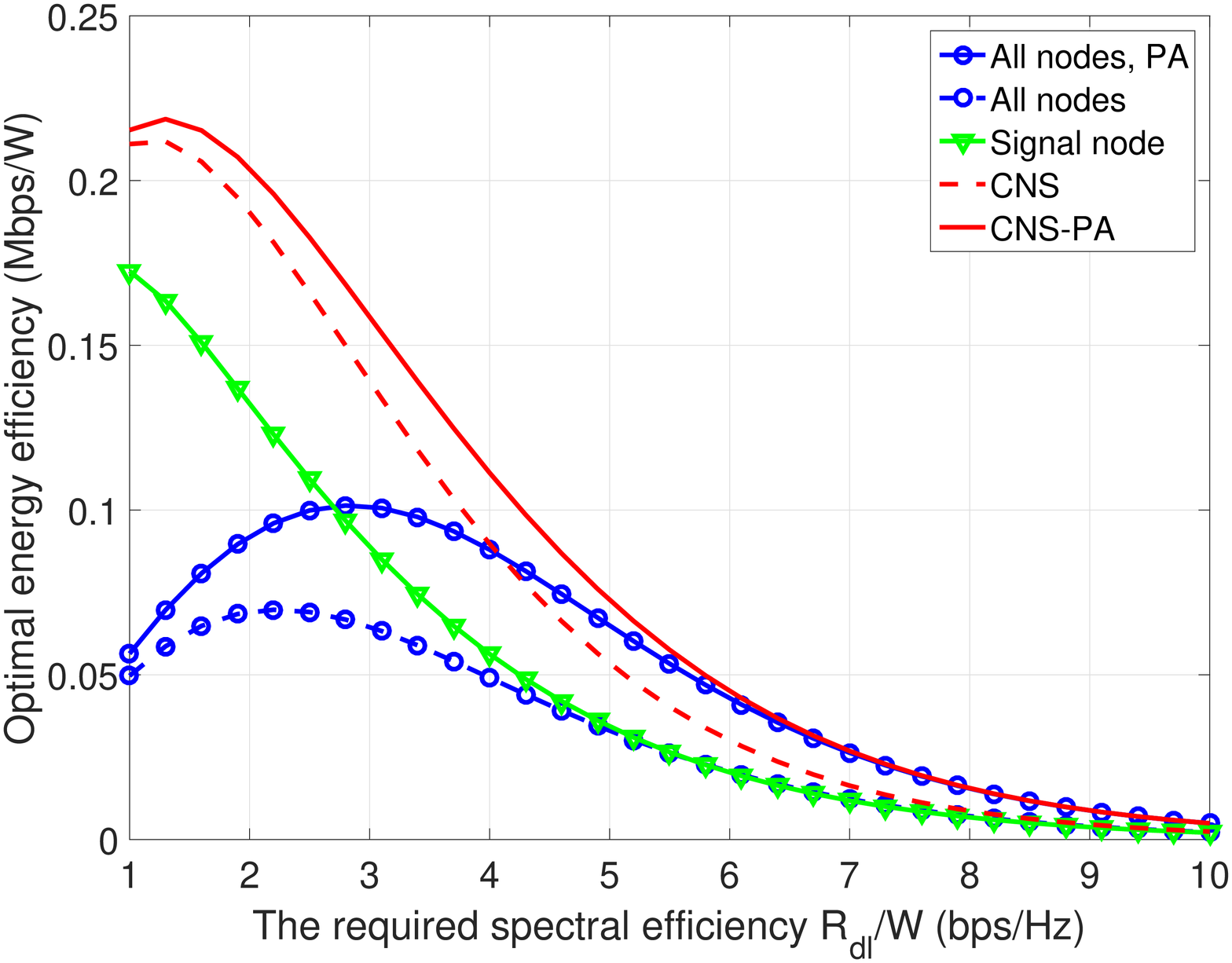}
\vspace{-6mm}
\caption{The optimal energy efficiency (Mbps/W) versus the required spectrum efficiency (bps/Hz) under ETPA with circuit power.}
\label{EEPerfprmance_ETPA}
\vspace{-5mm}
\end{figure}

\begin{figure}[!t]
\centering
\includegraphics[scale=0.25]{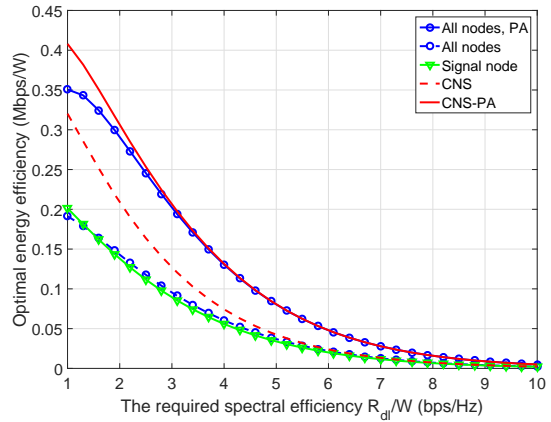}
\vspace{-6mm}
\caption{The optimal energy efficiency (Mbps/W) versus the required spectrum efficiency (bps/Hz) under IPA with circuit power.}
\label{EEPerfprmance_IPA}
\vspace{-5mm}
\end{figure}
Fig.\ref{EEPerfprmance_ETPA} and Fig.\ref{EEPerfprmance_IPA} compare the EE performance between ETPA and IPA with different required spectral efficiency $R_{\rm{dl}}/W$. It is clearly observed that our proposed CNS-PA scheme will attain optimal EE compared with other schemes both under ETPA and IPA. It also can be seen power allocation is a valid method to promote EE since the schemes with power allocation are all better than these without. Comparing the two subpictures, a conclusion is drawn that ETPA degrades the EE significantly due to much more extra energy consumed in power amplifier. Furthermore, the curves under ETPA first ascend in low data rate region and then decrease eventually, while under IPA these curves decrease directly without increment. The explanation is that the circuit power and extra power consumed by ETPA are independent with data rate and will deteriorate EE. Under lower data rate demand, these power consumptions dominate the total power consumption compared with transmission power and this part of EE will rise with the increase of data rate. However, when the data rate demand is high, transmission power will play the bigger role among total power consumption and EE generally decreases with the growth of data rate due to the exponentially increasing nature of transmission power with respect to data rate.

\begin{figure}[!t]
\centering
\includegraphics[scale=0.25]{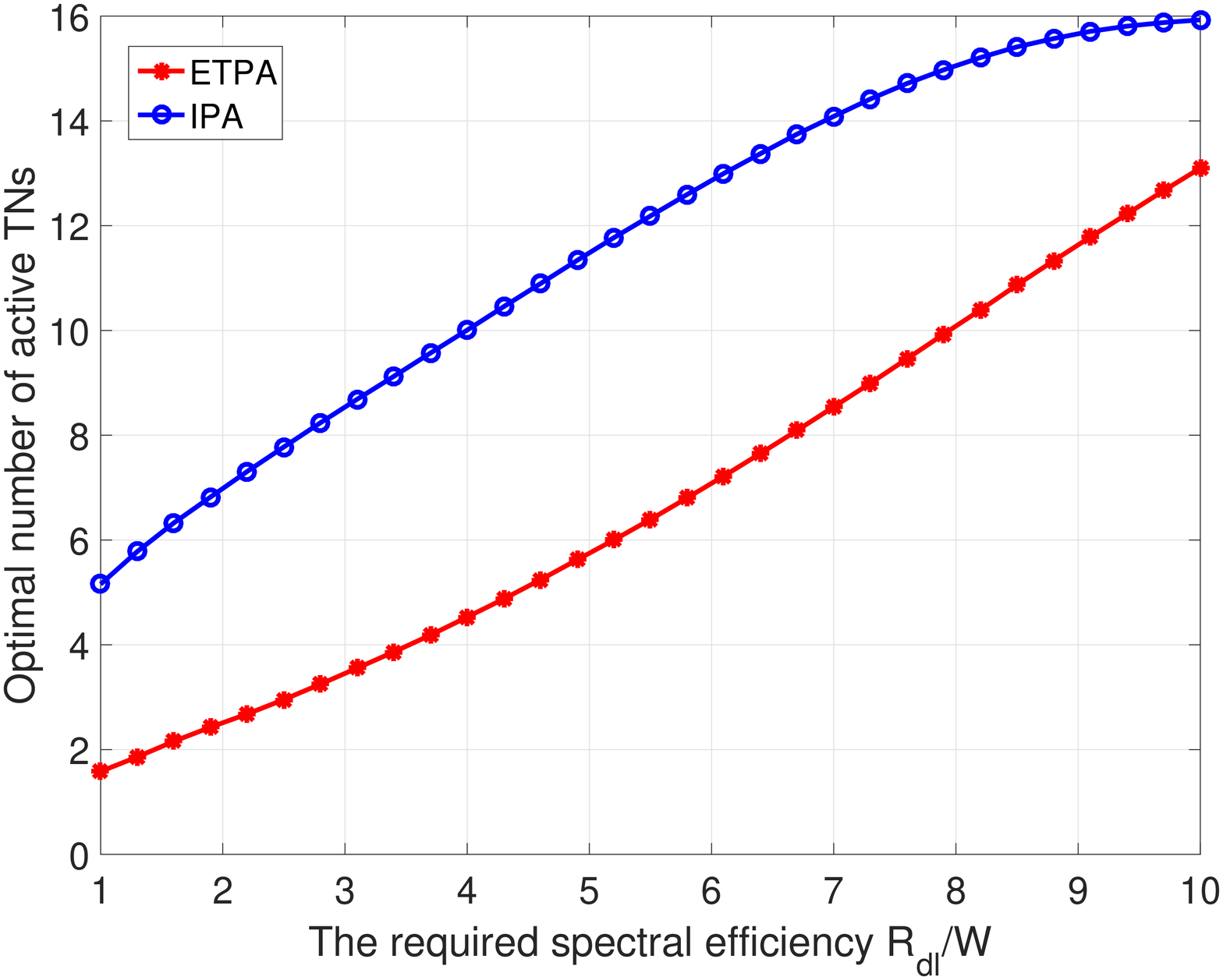}
\vspace{-6mm}
\caption{The optimal number of active TNs versus the required spectrum efficiency (bps/Hz) with circuit power in CNS-PA scheme.}
\label{ETPA_IPA_acvtiveNum}
\vspace{-5mm}
\end{figure}
Fig.\ref{ETPA_IPA_acvtiveNum} illustrates the optimal number of active TNs in CNS-PA scheme under ETPA and IPA, respectively. It is observed that more TNs will be selected with the increase of data rate requirement, which is because the transmission power will increase exponentially and it is much more larger than the circuit power and extra power consumed by ETPA. At this time, more TNs need to be involved to reduce transmission power even if it introduces more circuit power and extra ETPA power. And it is known that due to much more extra energy consumption introduced by ETPA, less TNs are involved under ETPA to reduce this part of power consumption, which is also the reason why the gap between CNS-PA and all nodes and PA scheme is bigger under ETPA in Fig.\ref{EEPerfprmance_ETPA}, especially in low data rate region.
\section{Conclusion}\label{Sec6}
In this paper, considering ETPA and circuit power, the energy-efficient CNS and power allocation scheme with data rate demand is proposed in coherent JT-CoMP networks. By establishing the selection priority lemma to reveal the optimal condition of CNS, the optimal cooperative TNs, whose powers can be calculated via the derived closed-form expressions, are selected according to the proposed selection criterion. Finally, Monte Carlo simulations are carried out to verify the superiority of our proposed CNS-PA scheme and show the effect of the circuit power and ETPA on the EE performance.
\section*{Acknowledgement}
The work was supported in part by National Nature Science Foundation of China Project under Grant 61471058, in part by the Key National Science Foundation of China under Grant 61461136002, in part by the Funds for Creative Research Groups of China under Grant 61421061, in part by the Hong Kong, Macao and Taiwan Science and Technology Cooperation Projects under Grant 2016YFE0122900, and in part by the 111 Project of China under Grant B16006​.

\end{document}